\documentclass[twocolumn,aps,prb,showpacs]{revtex4}
\usepackage{graphicx}
\usepackage{epsf}
\usepackage{amsmath}
\usepackage{amssymb}
\newcommand{\etal}{{\it et al.}}

\begin{document}

\title{Nodal lines and nodal loops in non-symmorphic odd-parity superconductors}

\author{T. Micklitz} 

\affiliation{
Centro Brasileiro de Pesquisas F\'isicas, Rua Xavier Sigaud 150, 22290-180, Rio de Janeiro, Brazil}

\author{M. R. Norman} 

\affiliation{Materials Science Division, Argonne National Laboratory,
  Argonne, Illinois 60439, USA}

\date{\today} 

\pacs{74.20.-z, 74.70.-b, 71.27.+a}

\begin{abstract}

We discuss the nodal structure of odd-parity superconductors in the presence of non-symmorphic crystal symmetries,
both with and without spin-orbit coupling, and with and without time reversal symmetry.  We comment on the relation
of our work to previous work in the literature, and also the implications for unconventional
superconductors such as UPt$_3$.

\end{abstract}
\maketitle

\section{Introduction}

Power-law temperature dependences of certain physical properties in heavy electron superconductors that were discovered
in the 1980s indicated the possible presence of nodes of the superconducting order parameter that form lines on the Fermi
surface.~\cite{pfleiderer} This motivated a study by Blount \cite{blount} where he showed that in the presence of
spin-orbit coupling, one would not expect line nodes for an odd-parity order parameter:  the
constraint of having all three components of the triplet vanish can happen at most at points on the Fermi surface.
This was an issue when an odd-parity order parameter was proposed to explain experimental data in UPt$_3$~\cite{sauls,norman}
which was consistent with later phase sensitive Josephson tunneling measurements.~\cite{strand}
In 1995, though, one of the authors found a possible solution to this problem, by showing that there is a counter example to Blount's theorem 
for non-symmorphic odd-parity superconductors (UPt$_3$ being such an example given its P6$_3$/mmc space group).~\cite{mike}
By explicit construction of the pair wave functions, it was found that on the zone face, $k_z=\pi/c$, all components of the triplet
belonged to the same group representation (as opposed to what happens on the $k_z=0$ zone plane), meaning that for
the proposed E$_{2u}$ symmetry, line nodes are indeed possible (two of the Fermi surfaces of UPt$_3$ intersect this zone face).
This was a consequence of the non-symmorphic phase factors associated with the c-axis (which is a screw axis for this space group). 
These considerations also potentially apply to other superconductors.  For instance, UBe$_{13}$ has the non-symmorphic
space group Fm$\bar{3}$c.

In 2009, a more rigorous treatment of this problem for the general non-symmorphic case was formulated by us based on group theoretical 
arguments.~\cite{NSLN}  Very recently, this problem has been revisited by Yanase \cite{yanase} and Kobayashi \etal.~\cite{sato} 
The former found that the nodal `lines' actually reconstruct to form nodal `loops' (called `rings' in the latter) which,
as we demonstrate here, shrink to zero as the ratio of the superconducting gap to the 
spin-orbit interaction increases. This is discussed in greater detail in Section III. The latter also discussed the mirror eigenvalues associated with
these nodal loops, as well as contrasted the group theoretical and topological approaches to this problem. 
It is our purpose here to clarify matters by a general group theoretical approach that in addition generalizes our previous work to
the case where spin-orbit interactions are absent, and also to the case of time reversal symmetry breaking.  We also consider the effect of glide-plane
symmetries, and find that these do not protect line nodes as do the screw-axis symmetries.

\section{Group theory}

The nodal structure of superconducting order parameters can be understood from representations  
of the symmetry group of the underlying crystal. The absence of certain representations on high-symmetry 
planes or lines in the Brillouin zone implies the presence of line or point nodes of the Cooper-pair 
wave function, respectively, in cases where the Fermi surface intersects these planes or lines. 
Representations of the superconducting order parameter in symmorphic crystals are readily 
found from the underlying point-group symmetries.~\cite{G-classification}
Non-symmorphic crystals, however, contain symmetries which 
consist of the combined operation of point-group elements with 
translations by fractions of a lattice vector. 
These non-primitive translations generate additional phase factors which have 
to be accounted for in the derivation of the 
Cooper-pair representations. Indeed, these phase factors 
may conspire in a way to exclude some of the symmetry-allowed representations 
on high symmetry planes, implying the possibility of new symmetry-enforced 
line nodes of the order parameter which are absent in symmorphic crystals.~\cite{mike,NSLN}
A convenient way to derive space-group representations of 
 the Cooper-pair wave function is to construct anti-symmetrized 
 products of the irreducible single-particle space-group representations,\cite{bradely,brad72,yarzhemsky}
 as we discuss next.

\subsection{Induced Cooper-pair representations} 

Consider a centrosymmetric crystal generated by a non-symmorphic space-group.
In the following, we denote space-group elements by $(g,\bold{t})$, where $g$ refers to the point-group operation 
and $\bold{t}$ accounts for possible non-primitive lattice translations, e.g.~$(I,0)$ is the inversion symmetry, etc.
Our focus here is on line nodes in odd-parity superconductors protected by non-symmorphic symmetries.  
We therefore concentrate on odd-parity representations of the Cooper-pair wave function at Brillouin-zone points $\bold{k}$ 
belonging to symmetry planes of non-symmorphic symmetry operations. 
Specifically, we consider 
symmetry planes $k_z=0, \pi$ of a glide-operation $(\sigma_z, \bold{t}_\sigma)$ 
and the combined action of inversion and two-fold screw axis, $(2_z, \bold{t}_2)(I,0)$ (from now on, we set the lattice constant to unity).
Here and in the following, $2_z$ denotes the two-fold rotation around the $z$-axis, 
$\sigma_z$ is reflection in the $z$-plane,
and $\bold{t}_{2/\sigma}$ is half a primitive translation along/perpendicular to 
the $z$-direction.
 
One can construct representations of the Cooper-pair wave function 
from the single-particle representations $\gamma_\bold{k}$
of symmetry-operations $m\in G_\bold{k}$ leaving $\bold{k}$ 
invariant (the ``little group").~\cite{brad72}  
To this end, one induces 
representations 
$P^-$ of the anti-symmetrized Kronecker-product with vanishing total momentum 
(modulo a reciprocal lattice vector) \cite{bradely,brad72,yarzhemsky} 
 \begin{align}
\label{eq:1}
\chi(P^-(m)) =& \chi(\gamma_\mathbf{k}(m)) 
\chi(\gamma_\mathbf{k}({\cal I} m {\cal I})), \\
\label{eq:2}
\chi(P^-({\cal I} m)) =& -\chi(\gamma_\mathbf{k}({\cal I} m {\cal I} m)),
\end{align} 
 where $\chi$ are the characters of the representation and 
 for notational convenience we introduced ${\cal I}\equiv(I,0)$.
In the presence of the spin-orbit interaction, the above equations  
characterize the pseudo-spin triplet components of the Cooper-pair wave function.
In the absence of spin-orbit, spin-rotational symmetry is conserved and they  
account for the Cooper pair's orbital degree of freedom of a spin-triplet state. 

Single-particle representations $\gamma_\bold{k}$ entering Eqs.~\eqref{eq:1}, \eqref{eq:2} 
are double- or single-valued, depending on the presence of the spin-orbit interaction.
Time-reversal symmetry $\theta$ can
 moreover induce extra degeneracies. These 
 are detected by Herring's criterion \cite{herring,lax} and taken into account by 
 passing to the corresponding co-representations (see Appendix \ref{app:1}).
We next apply the outlined procedure to construct 
Cooper-pair representations for the symmetries of interest.

\subsection{Two-fold screw axis}

\begin{table}[b!]
\begin{tabular}{p{1.2cm}|p{.8cm}p{2.1cm}p{.9cm}p{1.4cm}}
  & $(E,0)$ & \quad\,\,\,$(\sigma_z,\mathbf{t}_2)$  & $(I,0)$ & \quad $(2_z,\mathbf{t}_2)$ \\
\hline 
$k_z=\pi$ & $\quad d^2$ & $-\chi^2((\sigma_z,\bold{t}_2))$  & $-d$ & $\,\,\, \chi((\sigma_z^2,0))$
\\
$k_z=0$ & $\quad d^2$ & $\,\,\, \chi^2((\sigma_z,\bold{t}_2))$  & $-d$ & $-\chi((\sigma_z^2,0))$
 \end{tabular}
\caption{
Character table for
representations $P^-$ of anti-symmetrized Kronecker deltas induced  by single-particle representations 
of dimension $d$ on the high-symmetry planes. For notational convenience, we 
suppress $\gamma_\bold{k}$. 
}
\label{table:1}
\end{table} 
\begin{table}[b!]
\begin{tabular}{p{1cm}|p{1.2cm}p{1.2cm}p{1.2cm}p{1.2cm}}
 & $(E,0)$  & $(\sigma_z,\mathbf{t})$  & $(I,0)$ & $(2_z,\mathbf{t})$\\
\hline 
$A_g$ & $\quad 1$   & $\quad 1$ & $\quad 1$ & $\quad 1$  \\
$A_u$ & $\quad 1$   & $\,\,-1$ & $\,\,-1$ & $\quad 1$ \\ 
$B_g$ & $\quad 1$   & $\,\,-1$ & $\quad1$ & $\,\,-1$  \\
$B_u $ & $\quad 1$   & $\quad 1$ & $\,\,-1$ & $\,\, -1$  
\end{tabular}
\caption{Character table for the 
irreducible representations of the Cooper-pair wave function on high symmetry planes of 
a screw-axis/glide-plane ($\bold{t}=\bold{t}_{z/\sigma}$ 
for a screw-axis/glide-plane). The second column determines the mirror eigenvalue of the Cooper pair.}
\label{table:0}
\end{table}

\begin{table}[t!]
\begin{tabular}{p{.9cm}|p{.9cm}|p{1.4cm}|p{3.5cm}}
  SO &TRS  & BZ plane& irreducible components \\
\hline 
\hline 
yes & yes &$k_z=\pi$ & $P^-= A_g + 3 B_u$
\\
& &$k_z=0$ & $P^-= A_g + B_u + 2A_u$
\\
\hline 
yes & no  & $k_z=\pi$ & $P^-=  B_u $
\\
& & $k_z=0$ & $P^-= A_u$
\\
\hline 
no & yes & $k_z=\pi$ & $P^-= A_g +  B_u + 2A_u$
\\
& & $k_z=0$ & $P^-= B_u$
\\
\hline 
no & no  & $k_z=\pi$ & $P^-= A_u$
\\
&& $k_z=0$ & $P^-= B_u$
 \end{tabular}
\caption{
Decompositions of Cooper-pair representations into 
their irreducible components. Here $g$ and $u$ denote the even- and odd-parity 
representations and $A_g/B_u$ and $B_g/A_u$ are representations which are even and odd under reflection 
in the symmetry plane (i.e., mirror eigenvalues $\pm 1$), respectively. The results depend on the presence of 
time-reversal symmetry (TRS) 
and the spin-orbit interaction (SO). 
}
\label{table:3}
\end{table}

Consider first the presence of a two-fold screw symmetry $(2_z,\bold{t}_2)$  along the $z$-axis. 
Line nodes can be enforced on the symmetry planes $k_z=0, \pi$ characterized by the 
 little group $G_\bold{k}=\{(E,0),(\sigma_z,\bold{t}_2)\}$. Notice that in spite of 
its non-primitive translation vector $(\sigma_z,\bold{t}_2)=(2_z,\bold{t}_2){\cal I}$ is a symmorphic operation 
since the former can be removed by redefinition of the spatial origin (that is, it is a mirror plane, not a true glide plane). 
We next induce representations in the described manner, i.e.~by defining characters for the symmetry operations 
in $G_\bold{k}\cup {\cal I}G_\bold{k}=\{(E,0),(\sigma_z,\bold{t}_2),(I,0),(2_z,\bold{t}_2)\}$.

Recalling the multiplication rule for non-symmorphic group elements, \cite{brad72}
$(g_1,\bold{t}_1)(g_2,\bold{t}_2)=(g_1g_2,\bold{t}_1 + g_1\bold{t}_2)$, it is verified that ${\cal I}(\sigma_z,\bold{t}_2){\cal I}
=e^{-ik_z}(\sigma_z,\bold{t}_2)$. We can thus simplify characters in Eqs.~\eqref{eq:1},~\eqref{eq:2} 
for the symmetry planes of interest as summarized in Table~\ref{table:1}. 
From this table we then read off 
 irreducible components of the Cooper-pair representations given in Table~\ref{table:0}. 
Notice that the second column in Tables~\ref{table:1} and~\ref{table:0}
 determines the mirror eigenvalue of the Cooper pair. We are thus left with the task of
finding characters in the second and fourth column which  
depend on the underlying symmetries.

In the presence of the spin-orbit interaction, $\gamma_\bold{k}$ are double-valued 
with purely imaginary eigenvalues. That is,  
$\chi(\gamma_\bold{k}(\sigma_z^2,0))=-d$ and  $\chi(\gamma_\bold{k}(\sigma_z,\bold{t}_2))=\pm id$ 
with $d$ the dimension of $\gamma_\bold{k}$. 
Time-reversal symmetry may induce extra degeneracies.
Applying Herring's criterion one indeed detects   
(Kramers) degeneracies on both symmetry planes. That is, 
 $d=2$ and one has to consider the corresponding double-valued co-representations 
 (see Appendix~\ref{app:1} for details). 
If time-reversal symmetry is broken, $\gamma_\bold{k}$ are one-dimensional.
In the absence of the spin-orbit interaction, $\gamma_\bold{k}$ 
only account for the orbital degree of freedom, i.e.~are single-valued. 
That is, 
$\chi(\gamma_\bold{k}(\sigma_z^2,0))=d$ and $\chi(\gamma_\bold{k}(\sigma_z,\bold{t}_2))=\pm d$.
Herring's criterion then signals
degeneracies induced by time-reversal symmetry on the Brillouin zone face $k_z=\pi$. 
The latter are known as ``sticking of bands" induced by a two-fold screw axis~\cite{herring,lax,heine}, and one has 
to pass to the single-valued co-representation (see again Appendix~\ref{app:1} for details). 
When time-reversal symmetry is broken, $\gamma_\bold{k}$ are again one-dimensional.

All characters of the induced representations are summarized in Appendix~\ref{app:2}. 
Table~\ref{table:3} gives the decomposition of the resulting Cooper-pair representations 
into irreducible components of Table~\ref{table:0}.
The first four rows apply in the limit of a strong spin-orbit interaction. 
Following Anderson,\cite{anderson} analogues of Cooper-pair singlet and triplets
can then be constructed from Kramers degenerate states $\bold{k}$, $\theta I\bold{k}$
and their time-reversed partners $\theta \bold{k}$, $I\bold{k}$. 
The pseudo-spin singlet $d_0$ belongs to the one-dimensional even-parity representation $(g)$ 
and the pseudo-spin triplet states $d_x, d_y, d_z$ span
 the three-dimensional odd-parity representation $(u)$.\cite{pseudo-spin}
On the high symmetry planes, the representations are additionally characterized by their mirror eigenvalue, i.e.~pair-wave 
functions are even $(A_g,B_u)$ or odd $(B_g,A_u)$ under reflection about the plane.~\cite{foot1}

The first two rows show that transformation properties of pseudo-spin triplets with respect to the mirror plane 
change from the basal plane to the Brillouin zone face. That is,
in the presence of time-reversal symmetry,  
all possible pair representations are allowed on the basal plane $k_z=0$. 
This is in accordance 
with Blount's theorem, since $d_x$ and $d_y$ belong to one representation, and $d_z$ to the other.
On the Brillouin zone face, on the other hand, odd-parity representations 
which are odd under reflection in the plane 
are absent (that is, all components of $\bold{d}$ belong to the same representation).
This opens the possibility of symmetry protected line nodes when the Fermi surface intersects 
the Brillouin zone face and provides 
a counter example to Blount's theorem as previously discussed in Refs.~\onlinecite{mike} and \onlinecite{NSLN}. 
The third and fourth line describe situations in which Kramers degeneracy is lifted by strong 
time-reversal symmetry breaking. In this case, only one of the four Cooper pair functions survives, i.e.~the 
pseudo-spin triplet component formed from degenerate states $\bold{k}$, $I\bold{k}$
(we consider pairing of non-degenerate states later).
Time-reversal symmetry breaking thus opens the possibility of 
symmetry-protected line nodes on both symmetry planes. 
This has also been discussed in a recent work by Nomoto and Ikeda.~\cite{nomoto}

The last four rows apply in the absence of the spin-orbit interaction. The indicated 
representations then classify the orbital part of the pair wave function. This is  
combined with one of the three symmetric spin-triplet states to guarantee overall anti-symmetry 
of the pair wave function. In the absence of band degeneracies, representations are  
thus one-dimensional, as in the last three rows, allowing for symmetry protected line nodes
on both symmetry planes.
In the presence of time-reversal symmetry, the two-fold screw axis induces, however, 
sticking of bands on the Brillouin zone face~\cite{herring,lax,heine} (fifth row).
One thus finds four allowed 
representations and both mirror eigenvalues are realized.
In the absence of both the spin-orbit interaction and time reversal breaking, 
symmetry-protected line-nodes 
are thus possible on the basal plane but 
do not exist on the Brillouin zone face.
The difference from the first two lines of this Table is that this sticking of bands allows the formation of interband
pairs in this case.~\cite{yanase}
The interband pairs are odd in the band index,
implying that the intra-orbital part of the Cooper pair wave function is even 
to guarantee overall odd parity (that is, they have opposite mirror eigenvalues to the intraband pairs).
We will return to this point below.  
In the absence of time-reversal symmetry, protected line nodes can appear on both symmetry planes, 
independent of the spin-orbit interaction. 
Finally, we note that for time reversal symmetry breaking, the inversion of the sign of the Cooper-pair mirror 
eigenvalue of one-dimensional representations in the presence (third and fourth rows) and absence 
(seventh and eighth rows) of the spin-orbit interaction
is readily related to the double- and single-valuedness of the representations.

\subsection{Glide plane}

\begin{table}[b!]
\begin{tabular}{p{1.3cm}|p{1cm}p{2.6cm}p{.8cm}p{1.4cm}}
  & $(E,0)$ & \quad\,\,\,$(\sigma_z,\mathbf{t}_\sigma)$  & $(I,0)$ & \quad $(2_z,\mathbf{t}_\sigma)$ \\
\hline 
$k_z=\pi,0$ & $\quad d^2$ & $e^{-ik_x}\chi^2((\sigma_z,\bold{t}_\sigma))$  & $-d$ & $- \chi((\sigma_z^2,0))$
 \end{tabular}
\caption{
Character table for
representations $P^-$ of anti-symmetrized Kronecker deltas induced  by single-particle representations 
of dimension $d$ on the high-symmetry planes. 
 For notational convenience, we 
suppress $\gamma_\bold{k}$, and 
assume $\bold{t}_\sigma$ parallel to the $x$-axis. 
}
\label{table:4}
\end{table} 
\begin{table}[t!]
\begin{tabular}{p{.9cm}|p{1.1cm}|p{1.4cm}|p{3.4cm}}
 SO & TRS &BZ plane& irreducible components \\
\hline 
\hline
  yes& yes &$k_z=\pi,0$ & $P^-= A_g + B_u + 2A_u$
\\
\hline 
yes & no & $k_z=\pi,0$ & $P^-=  A_u $
\\
\hline 
no & yes/no & $k_z=\pi,0$ & $P^-= B_u$
\end{tabular}
\caption{
Decompositions of Cooper-pair representations into their
irreducible components. The latter depend on the presence of 
time-reversal symmetry (TRS) 
and the spin-orbit interaction (SO). Here,
$g/u$ denote representations which 
are even/odd under inversion and $A_g/B_g$, respectively, $B_u/A_u$  which 
are even/odd under reflection 
in the symmetry plane. 
}
\label{table:6}
\end{table} 

Consider next a glide-plane symmetry $(\sigma_z,\bold{t}_\sigma)$,
where without loss of generality we can assume $\bold{t}_\sigma$ parallel to the $x$-axis. 
The little group on the symmetry planes $k_z=0, \pi$ is $G_\bold{k}=\{(E,0),(\sigma_z,\bold{t}_\sigma)\}$ 
and we induce representations for the symmetry operations in  
$G_\bold{k}\cup {\cal I}G_\bold{k}=\{(E,0),(\sigma_z,\bold{t}_\sigma),(I,0),(2_z,\bold{t}_\sigma)\}$. 
Here $(2_z,\bold{t}_\sigma)$ in spite of its non-primitive translation is a symmorphic operation 
(again the translation can be removed by redefinition of the spatial origin). 
 Using the commutation relation ${\cal I}(\sigma_z , \bold{t}_\sigma){\cal I} = e^{-ik_x}(\sigma_z, \bold{t}_\sigma)$,
 characters of the induced representations can be simplified, as shown in Table~\ref{table:4}.  
 The induced representations are identical on both symmetry planes.  
 The dimension $d$ and characters for $(\sigma_z,\bold{t}_\sigma)$ and $(\sigma_z^2,0)$ 
depend again on the underlying symmetries.

Let us first consider the presence of the spin-orbit interaction
with double-valued representations, $\chi(\gamma_\bold{k}(\sigma_z^2,0))=-d$. 
If time-reversal symmetry is preserved, 
Herring's criterion indicates the presence of
(Kramers) degeneracies on both symmetry planes. That is, $d=2$ and 
we need to pass  
to the double-valued co-representation (see Appendix~\ref{app:1} for details).
If time-reversal symmetry is broken, $\gamma_\bold{k}$ remain one-dimensional 
and $\chi(\gamma_\bold{k}(\sigma_z,\bold{t}_\sigma))=\pm ie^{ik_x/2}$. 
In the absence of the spin-orbit interaction, on the other hand, all single-particle representations 
are one-dimensional, independent of 
time-reversal symmetry.

All characters of the induced representations are summarized in Appendix~\ref{app:2}. 
Table~\ref{table:6} shows the decompositions of the resulting  Cooper-pair representations into 
irreducible components of Table~\ref{table:0}.
If the spin-orbit interaction and time-reversal symmetry are both present,
all odd-parity representations are allowed on both planes.
That is, glide-plane symmetries do not provide us with counter examples to Blount's theorem.
In the absence of either time-reversal symmetry or the spin-orbit interaction, symmetry-protected line nodes can occur 
on both symmetry planes.

Our discussion 
so far has shown that in the presence of time-reversal symmetry and the spin-orbit interaction, only two-fold screw 
axes  can protect line nodes in odd-parity superconductors. 
Next, we discuss that these line nodes typically form as loops.

\section{Nodal structure of odd-parity superconductors}

\begin{figure}
\includegraphics[width=\hsize]{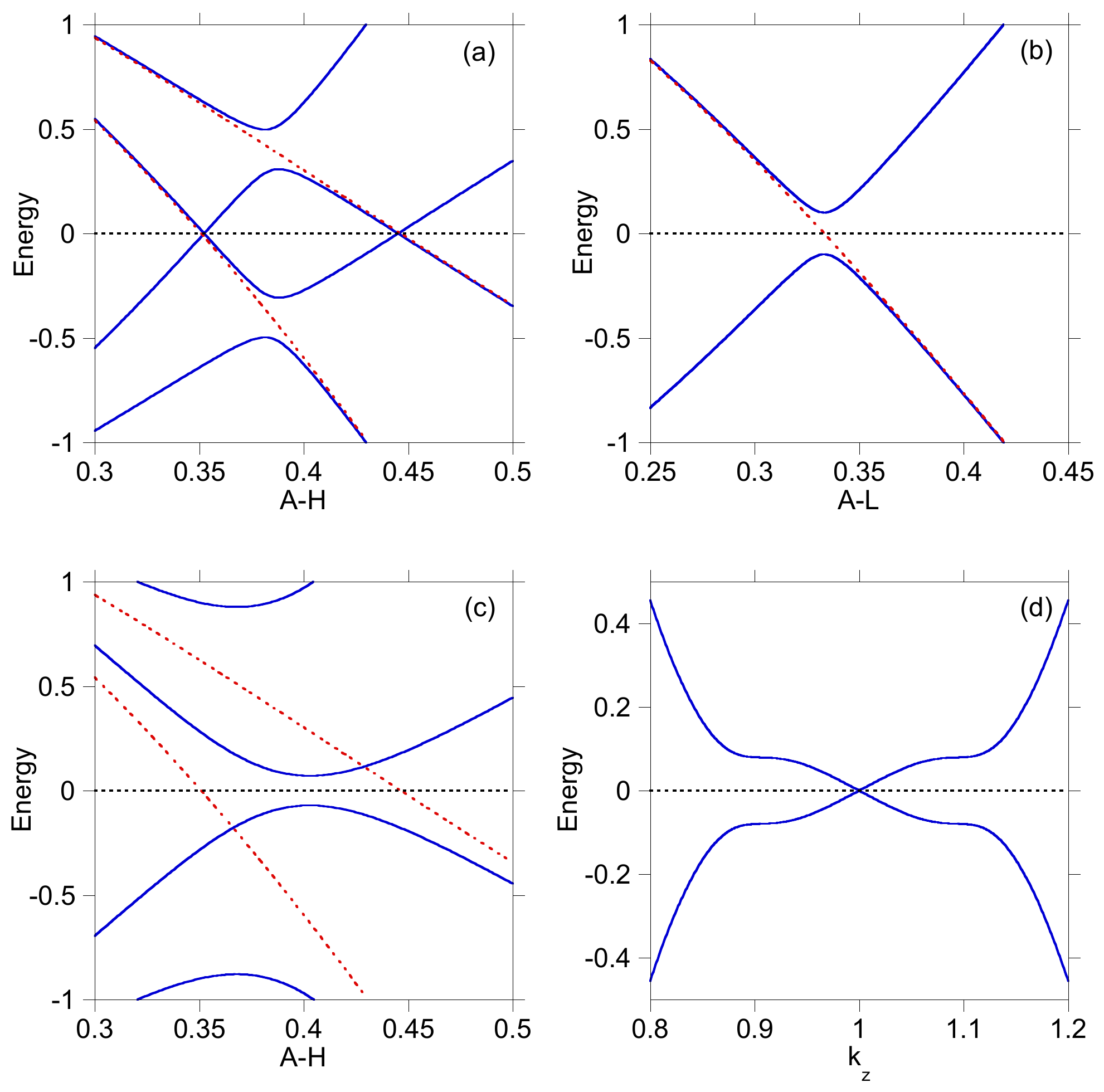}
\caption{Illustration of nodal loops in UPt$_3$ based on the toy model of Yanase.~\cite{yanase}  Solid curves indicate the superconducting state
dispersion, dashed curves the normal state dispersion.  (a) dispersion along $A$-$H$ for $\Delta$=0.1, where $A$ is (0,0,$\pi/c$) and $H$ is (0,4/3$a$,$\pi/c$).
(b) dispersion along $A$-$L$ for $\Delta$=0.1, where $A$ is (0,0,$\pi/c$) and $L$ is (2$\pi$/$\sqrt{3}a$,0,$\pi/c$).
Here, $\Delta$ is the value of the superconducting E$_{2u}$ order parameter in these energy units, this being the $f$ function of Yanase
which pairs electrons between two near-neighbor uranium sites (taken here as a constant for illustrative purposes).
The nodes in (a) (due to the absence of intraband pairing for E$_{2u}$ symmetry) 
and their lack thereof in (b) (due to interband pairing, which is allowed for this symmetry) lead to the two nodal lines closing to form nodal loops
in the $k_z$=$\pi/c$ zone face.
(c) Same as (a), but for $\Delta$=0.5, showing the disappearance of the nodes along $A$-$H$, and thus the collapse of the nodal loops.\cite{foot2}
(d) dispersion along $k_z$ normal to the second node along $A$-$H$ in (a), illustrating that these are nodal loops, and not toroidal Fermi surfaces.
This can also be seen from plots like in (a), where the nodes lift when k$_z$ deviates from $\pi/c$.}
\label{fig1}
\end{figure}

\begin{figure}
\includegraphics[width=0.7\hsize]{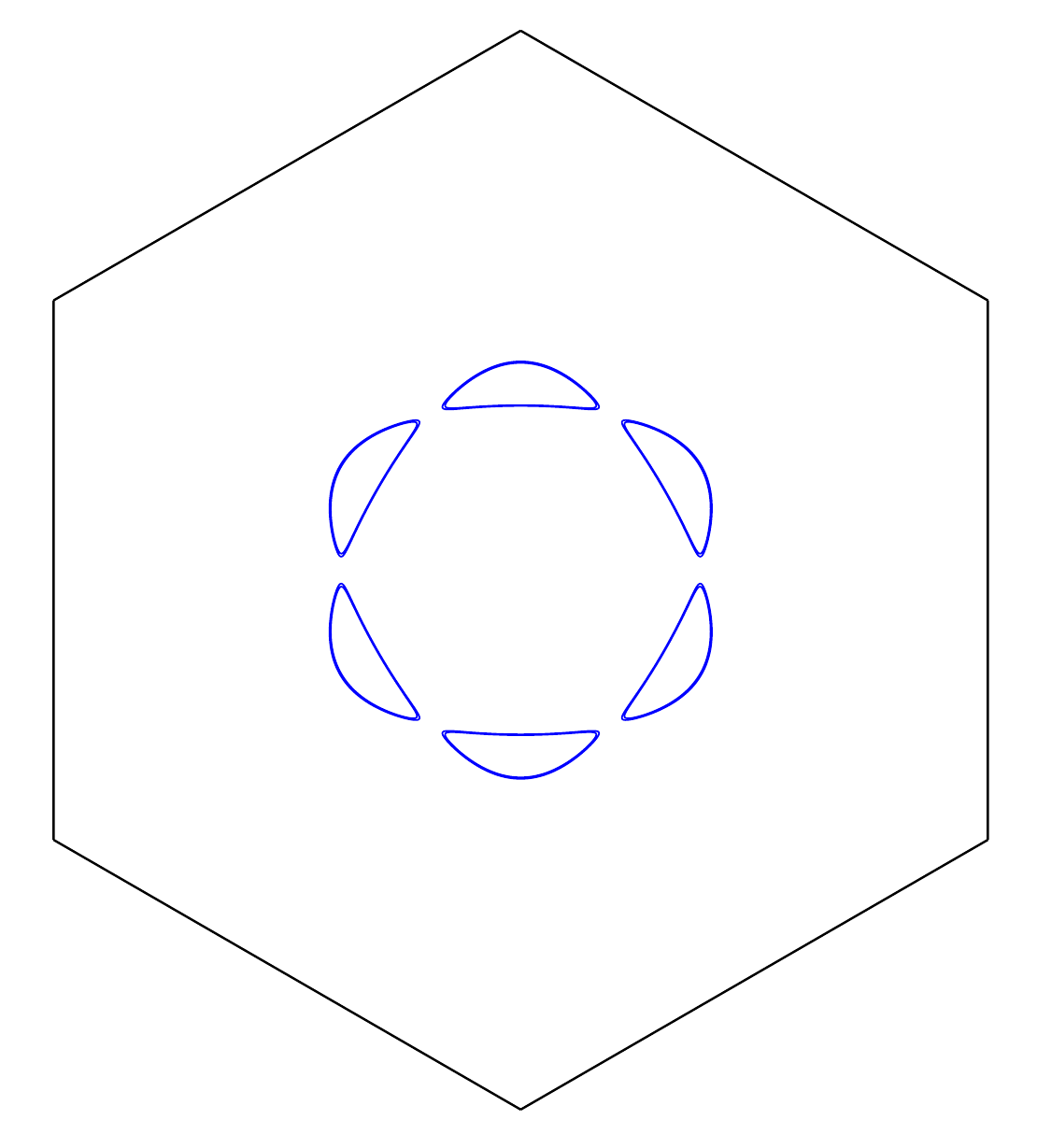}
\caption{Nodal loops in the $k_z$=$\pi/c$ zone face using the parameters from Fig.~1.  They form due to the energy gap from interband
pairing that occurs along the $A$-$L$ lines.}
\label{fig2}
\end{figure}

As pointed out by Yanase,\cite{yanase} the nodal lines discussed above in the $k_z=\pi$ zone face actually
reconstruct to form nodal loops in the case of UPt$_3$.
The latter are, in contrast to line nodes, contractible, i.e.~they continuously shrink to zero as the ratio of 
the superconducting 
gap to the spin-orbit interaction increases. 
The formation of these nodal loops can be understood from the results of Section II.
In particular, along the $A$-$L$ lines of this zone face, the spin-orbit interaction vanishes, leading to band sticking
(this band sticking effect has been seen in UPt$_3$ from breakdown orbits in deHaas-vanAlphen
measurements \cite{greg}).  This means that interband pairs can form at these sticking points on the Fermi surface,
and since they have opposite mirror eigenvalues, they are allowed representations for the case where the
intraband pairs are not allowed.  This leads to a gapping of the Fermi surface at these points, thus converting the 
nodal lines to nodal loops, as we illustrate in Figs.~1 and 2. 
As the order parameter increases, these nodal loops will eventually shrink to zero, leading to a topological
transition (Fig.~1c).  In Ref.~\onlinecite{sato}, topological arguments are presented (following earlier work \cite{koba})
that confirm the group theoretical ones.
There, a claim was made that the topological arguments are more general than the group theory ones,
but in fact they are equivalent.  In particular, as we showed
in Section II, in the presence of time reversal symmetry breaking, the nodal structure of the pairs changes due
to lifting of the degeneracy of the single-particle states.

Although much of the discussion above was motivated by UPt$_3$, there are other superconductors that
have non-symmorphic space groups. 
We earlier mentioned UBe$_{13}$.  But its space group does not have a screw axis, but rather a glide plane, so
we would not expect nodal lines in this case for odd parity pairing, which is consistent with specific heat data.~\cite{ott} 
But, URhGe, UCoGe, and UIr have screw axes, though the last breaks inversion symmetry,
meaning even and odd parity can mix.~\cite{pfleiderer}
Moreover, the presence of magnetism can induce
non-symmorphic behavior and nodal lines as recently discussed in the context of UCoGe and UPd$_2$Al$_3$.~\cite{nomoto}
Yet to be explored are the consequences of the effects discussed here on potential topological surface states. This 
will be addressed in future work.

\begin{acknowledgments}

This work was supported by the Materials Sciences and Engineering
Division, Basic Energy Sciences, Office of Science, US Dept.~of Energy.
T. M. acknowledges financial support by Brazilian agencies CNPq and FAPERJ.

\end{acknowledgments}

\begin{appendix}

\section{Herring's criterion and co-representations}
\label{app:1}

As stated in the main text,
time-reversal symmetry $\theta$ can
induce additional degeneracies. In this case, one should pass from the representations 
to corresponding co-representations of the magnetic group
${\cal G}_\bold{k}=G_\bold{k}+ {\cal I}\theta G_\bold{k}$. 
Degeneracies induced by $\theta$ can be detected by Herring's criterion
from the sum of characters, \cite{lax}
\begin{align}
\label{eq:3}
\sum_{ B\in  G_\bold{k}} \chi\left(\gamma_\bold{k}({\cal I}\theta B)^2\right) 
=\begin{cases} 
+ |G_\bold{k}| & \quad \text{case (a)}\\
- |G_\bold{k}| & \quad \text{case (b)}\\
\quad 0 & \quad \text{case (c).}
\end{cases}
\end{align} 
Here $|G_\bold{k}|$ is the order of the little group.
In case (a) no degeneracies are induced, 
while (b) and (c) indicate the presence of degeneracies. The latter    
are accounted for by passing to co-representations 
 $\gamma_\bold{k} \mapsto \Gamma_\bold{k} \equiv \left(\begin{smallmatrix} 
 \gamma_\bold{k} & \\ & \bar{\gamma}_\bold{k} \end{smallmatrix}\right)$,
 where $\bar{\gamma}_\bold{k}(m)=\gamma_\bold{k}(m)$ in case (b)  and
$\bar{\gamma}_\bold{k}(m)=\gamma_\bold{k}^*(({\cal I} \theta )^{-1}\, m \, {\cal I}\theta)$ in case (c), respectively, 
with `\,$^\ast$\,' the complex conjugation. 
We next consider the cases of interest.

{\it Two-fold screw axis:---} In presence of the spin-orbit interaction, 
the sum of characters for double-valued representations reads 
\begin{align}
&\chi\left([{\cal I}\theta (E,0)]^2\right)
+
\chi\left([{\cal I}\theta (\sigma_z,\bold{t}_2)]^2\right)
\nonumber\\
&=
- \chi\left((E,0)\right)
-
e^{ik_z}
\chi\left((\sigma_z^2,0)\right)
\nonumber\\
&=
-1+e^{ikz},
\end{align}
where we used that  $\theta g_1 \theta g_2=-g_1g_2$.
For double valued co-representations on the basal-plane (case (c)), we then employ  
$\bar{\gamma}_\bold{k}((\sigma_z,\bold{t}_2))
=
\gamma_\bold{k}^*\left(({\cal I}\theta)^{-1} (\sigma_z,\bold{t}_2)\,{\cal I}\theta\right)
= \gamma_\bold{k}^*((\sigma_z,\bold{t}_2))$. 
For single-valued representations, on the other hand,
$\theta g_1 \theta g_2=g_1g_2$ and 
\begin{align}
&\chi\left([{\cal I}\theta (E,0)]^2\right)
+
\chi\left([{\cal I}\theta (\sigma_z,\bold{t}_2)]^2\right)
\nonumber\\
&=
\chi\left((E,0)\right)
+
e^{ik_z}\chi\left((\sigma_z^2,0)\right)
\nonumber\\
&=
1
+
e^{ik_z}.
\end{align}
For the single-valued co-representation on the Brillouin zone face (case (c)), we use that  
$\bar{\gamma}_\bold{k}((\sigma_z,\bold{t}_2))
=
\gamma_\bold{k}^*\left(({\cal I}\theta)^{-1} (\sigma_z,\bold{t}_2)\,{\cal I}\theta\right)
=- \gamma^*_\bold{k}\left((\sigma_z,\bold{t}_2)\right)$.

{\it Glide-plane symmetry:---} In the presence of the spin-orbit interaction
\begin{align}
&\chi\left([{\cal I}\theta (E,0)]^2\right)
+
\chi\left([{\cal I}\theta (\sigma_z,\bold{t}_\sigma)]^2\right)
\nonumber\\
&=
- \chi((E,0))
+
\chi\left((\sigma_z^2,0)\right)
\nonumber\\
&=
0,
\end{align}
and for the double valued co-representations (case (c)), we then employ  
$\bar{\gamma}_\bold{k}((\sigma_z,\bold{t}_\sigma))
=
\gamma_\bold{k}^*\left(({\cal I}\theta)^{-1} (\sigma_z,\bold{t}_z)\,{\cal I}\theta\right)
= e^{ik_x}\gamma^*_\bold{k}((\sigma_z,\bold{t}_2))$,
i.e.~$\Gamma_\bold{k}((\sigma_z,\bold{t}_\sigma))
=
\pm e^{ik_x/2}
(\begin{smallmatrix}
i &
\\
&
- i
\end{smallmatrix})$. 
In the absence of the spin-orbit interaction 
\begin{align}
&\chi\left([{\cal I}\theta (E,0)]^2\right)
+
\chi\left([{\cal I}\theta (\sigma_z,\bold{t}_\sigma)]^2\right)
\nonumber\\
&=
\chi\left((E,0)\right)
+
\chi\left((\sigma_z^2,0)\right)
\nonumber\\
&=
2.
\end{align}

\section{Irreducible representations of the Cooper-pair wave function}
\label{app:2}

We summarize the characters of induced representations in the case of a two-fold screw axis (Table~\ref{table:2})
and a glide-plane symmetry (Table~\ref{table:5}).
The decompositions of Cooper-pair representations into  
their irreducible components are done  using the character table
for  
the zero-momentum representations of the Cooper-pair wave function defined in Table~\ref{table:0} 
in the main text.

\begin{table}[h!]
\begin{tabular}{p{.9cm}|p{1.1cm}|p{1.4cm}|p{.8cm}p{1cm}p{.7cm}p{.9cm}}
  SO & TRS &BZ plane& $(E,0)$ & $(\sigma_z,\mathbf{t}_2)$  & $(I,0)$ & $(2_z,\mathbf{t}_2)$ \\
\hline
\hline 
yes & yes &$k_z=\pi$ & \quad$4$ & \quad$\,4$  & $-2$ &$\,\,-2$
\\
& &$k_z=0$ & \quad$4$ & \quad\,$0$  & $-2$ &\quad $\,2$
\\
\hline 
yes & no &$k_z=\pi$ & \quad$1$ & \quad$\, 1$  & $-1$ & $\,\,-1$
\\
& & $k_z=0$ & \quad$1$ & $\,\,-1$  & $-1$ & \quad$\,1$
\\
\hline 
no & yes & $k_z=\pi$ & \quad$4$ & \quad$\,0$  & $-2$ & \quad$\,2$
\\
& & $k_z=0$ &\quad $1$ &  \quad\,$1$  &$-1$ & $\,\,-1$
 \\
\hline 
no & no &$k_z=\pi$ & \quad$1$ & $\,\, -1$  & $-1$ & \quad\,$1$
\\
& & $k_z=0$ & \quad$1$ & \quad$\,1$  & $-1$ & $\,\,\,-1$
 \end{tabular}
\caption{
Character table for
representations $P^-$ of anti-symmetrized Kronecker deltas on symmetry planes
induced by single-particle representations. 
Depending on the presence of time-reversal symmetry (TRS) 
and the spin-orbit interaction (SO),
the latter are single or double-valued (co-)representations.
}
\label{table:2}
\end{table}

\begin{table}[h!]
\begin{tabular}{p{.9cm}p{1.1cm}|p{1.4cm}|p{.8cm}p{1cm}p{.7cm}p{.9cm}}
  SO & TRS &BZ plane& $(E,0)$ & $(\sigma_z,\mathbf{t}_\sigma)$  & $(I,0)$ & $(2_z,\mathbf{t}_\sigma)$ \\
\hline 
\hline 
yes & yes &$k_z=\pi,0$ & \quad$4$ & \quad$\,0$  & $-2$ &$\quad \, 2$
\\
\hline 
yes & no &$k_z=\pi,0$ & \quad$1$ & \quad$-1$  & $-1$ & $\quad\,1$
\\
\hline 
no & yes/no & $k_z=\pi,0$ & \quad$1$ & \quad$\, 1$  & $-1$ & \quad$-1$
\end{tabular}
\caption{
Character table for
representations $P^-$ of anti-symmetrized Kronecker deltas on symmetry planes
induced by single- or double-valued (co-)representations.
}
\label{table:5}
\end{table}

\end{appendix}


\begin{thebibliography}{99}
\bibitem{pseudo-spin}
The pseudo-spin singlet (corresponding to $S=0$ in the absence of the spin-orbit interaction)
is given by the even-parity combination $(\bold{k},\theta\bold{k})-(\theta I \bold{k},I\bold{k})$. 
The three pseudo-spin triplet components (corresponding to $S=1$ in the absence of the spin-orbit interaction) 
are given by the odd-parity combinations $(\bold{k}, I\bold{k})$, $(\theta I \bold{k}, \theta \bold{k})$, 
and $(\bold{k}, \theta \bold{k})+(\theta I \bold{k}, I \bold{k})$. 
The latter are conveniently relabeled as a vector $\bold{d}$ and the above three states correspond 
to the components $-d_x +id_y$, $d_x +id_y$, and $d_z$, respectively.
\bibitem{foot1}
Our notation for $A_u$ and $B_u$ agrees with previous literature \cite{yip} that takes into account 
that $\bold{d}$ transforms as an axial vector.
\bibitem{foot2}
Calculations were done with the same parameters as in Ref.~\onlinecite{yanase}, that is
$t$=1, $t_z$=-4, $t_p$=1, $\alpha$=2, and $\mu$=12.  The critical value of $\Delta$ for the nodal loop
to disappear is given by $\Delta_{cr}$=$\sqrt{-\epsilon_+ \epsilon_-}$ where $\epsilon_{\pm}$ = $\epsilon_k \pm g_k$
with $\epsilon_k$ the normal state dispersion (without spin-orbit), $g_k$ the spin-orbit splitting,
and $k$ the value of the momentum at the dispersion minimum seen in (c).
\end{thebibliography}
\end{document}